\documentstyle[prd,aps]{revtex}

\newcommand{\bea}{\begin{eqnarray}}
\newcommand{\eea}{\end{eqnarray}}

\begin{document}

\draft
\twocolumn[\hsize\textwidth\columnwidth\hsize\csname
@twocolumnfalse\endcsname

\title{Cosmological Perturbations with Multiple Scalar Fields}
\author{Jai-chan Hwang${}^{(a)}$ and
        Hyerim Noh${}^{(b)}$ \\
        \small ${}^{(a)}$ Department of Astronomy and Atmospheric Sciences, 
                   Kyungpook National University, Korea \\
        \small ${}^{(b)}$ Korea Astronomy Observatory 
                   Whaam-dong, Yusung-gu, Korea}
\date{\today}
\maketitle

\begin{abstract}

We consider the evolution of perturbed cosmological spacetime with multiple 
scalar fields in Einstein gravity.
A complete set of scalar-type perturbation equations is presented in
a gauge-ready form, and we derived the closed set of second-order 
differential equations in several useful forms.
Conserved behaviors of the perturbed three-space curvature in the comoving 
gauge, $\varphi_v$, under several conditions are clarified.
Under the slow-roll conditions, the adiabatic and isocurvature modes
decouple from each other, and in the large-scale limit we have
(i) the adiabatic mode is generally conserved,
(ii) for a couple of special potential the isocurvature modes decouple 
from each other and are described by conserved quantities, 
(iii) in the two field system, the isocurvature mode is described by 
a conserved quantity for the general potential.

\end{abstract}


\vskip2pc]

\section{Introduction}
                                          \label{sec:Introduction}

The cosmological perturbation theory \cite{Lifshitz-1946}, 
handling the evolution of linearly
perturbed spacetime and matter in otherwise spatially homogeneous
and isotropic world model, plays central role in the modern 
theory of large-scale structure formation in the universe.
The linearity assumption of the structures seems to hold 
in the early evolution stage and in the large scale evolution of
our presently observable patch of the universe.
Besides the quantitative constraint on the cosmological 
model implying the flat universe, the recent observations of the CMBR 
anisotropies in the small-angular scale by Boomerang and Maxima-1 experiments 
\cite{deBernardis-2000}, in our opinion, confirm dramatically the validity 
of the basic assumptions used in the cosmological perturbation theory, 
i.e., the linearity of the relevant cosmic structures in such scales.

The theoretical introduction of an acceleration phase in the
early universe \cite{Inflation} 
enables us to draw a coherent picture of the structure generation process.
The ever-present microscopic vacuum quantum fluctuations
inside a causal domain can rapidly expand to become
macroscopic during this acceleration phase, which
in turn, can evolve into large-scale structures in the universe.
The acceleration phase is often modelled by using a scalar field
with a relevant potential term. 
Thus, while the background scalar field drives the early universe accelerating,
the quantum fluctuation of the scalar field accompanied with 
simultaneously excited spacetime metric fluctuations provides
seeds for the large-scale structure and the gravitational wave.
Recent development shows the possibility that observational constraints 
can be translated into the constraints on, or the reconstructions of, 
the form of field potential driving the inflation.
Such a reconstruction program is popular in the context of an inflation 
driven by a single minimally coupled scalar field with the slow-roll 
conditions \cite{Reconstruction}.

The early universe scenarios with multiple episodes of inflation
have recently attracted much attention in the literature.
Such scenarios apparently allow more freedom in designing the spectra of 
the large-scale cosmic structures in their generation and evolution stages
\cite{Salopek-etal-1989}.
{}For the benefit of future studies, in this paper we would like to 
present the perturbed scalar field equations valid for an arbitrary number
of scalar fields in the gauge-ready form, derive closed form equations
in several useful forms, and derive general solutions in the large-scale limit.
We will find that in various generic situations the evolution of perturbations
can be characterized by certain conserved quantities: the adiabatic 
perturbation characterized by the perturbed curvature variable in the
comoving gauge, $\varphi_v$, is conserved in various situations, and the
isocurvature perturbations defined by a relative perturbation among
perturbed fields, $\delta \phi_{IJ}$, are often described by some
conserved quantities, as well.
We start by introducing our notations, general equations, and strategy
in \S \ref{sec:background}.
We set $c \equiv 1$.

\section{Perturbed world model and basic equations}
                                          \label{sec:background}

We consider a gravity with an arbitrary number of scalar fields 
minimally coupled with the gravity.
As the Lagrangian we consider
\bea
   L = {1 \over 16 \pi G} R
       - {1\over 2} \sum_J \phi^{J;c} \phi_{J,c} - V (\phi^K).
   \label{Lagrangian}
\eea
$R$ is the scalar curvature, and $a,b, \dots$ are spacetime indices.
$\phi^I$ is the $I$-th component of $N$ scalar fields with 
$I, J, \dots = 1,2, \dots, N$; $I,J$ indices are raised and lowered 
by $\delta_{IJ}$.
$V(\phi^K)$ is a general algebraic function of 
the scalar fields, i.e., $V(\phi^K) = V (\phi^1, \dots , \phi^N)$.
Variations with respect to $g_{ab}$ and $\phi^I$ lead to
the gravitational field equation and the equation of motion:
\bea
   & & G_{ab} = 8 \pi G \left[ \sum_J \left( \phi^J_{\;\;,a} \phi_{J,b}
       - {1 \over 2} g_{ab} \phi^{J;c} \phi_{J,c} \right) - V g_{ab} \right]
   \nonumber \\
   & & \quad
       \equiv 8 \pi G T_{ab},
   \label{GFE} \\
   & & \Box \phi^I - V^{,I} = 0,
   \label{EOM}
\eea
where $V_{,I} \equiv \partial V/(\partial \phi^I)$.

We consider the general perturbations in the FLRW world model.
In the spatially homogeneous and isotropic background we are considering,
to the linear order the three different types of perturbation
decouple from each other and evolve independently.
The presence of multiple number of minimally coupled scalar fields
does {\it not affect} the graviational wave of the perturbed spacetime 
and the the rotational perturbations of the additional fluids, directly.
Thus, in the following we will consider the {\it scalar-type} perturbation only.
As the metric we take
\bea
   & & d s^2 = - a^2 \left( 1 + 2 \alpha \right) d \eta^2 
       - a^2 \beta_{,\alpha} d \eta d x^\alpha
   \nonumber \\
   & & \quad
       + a^2 \left[ g^{(3)}_{\alpha\beta} \left( 1 + 2 \varphi \right)
       + 2 \gamma_{,\alpha|\beta} \right] d x^\alpha d x^\beta,
   \label{metric-general}
\eea
where $a(\eta)$ is the cosmic scale factor.
$\alpha$, $\beta$, $\gamma$ and $\varphi$
are spacetime dependent perturbed order variables.
A vertical bar $|$ indicates a covariant derivative based on 
$g^{(3)}_{\alpha\beta}$.
Considering the FLRW background and the scalar-type perturbation,
the energy-momentum tensor can be decomposed into the effective fluid 
quantities as
\bea
   & & T^0_0 = - \left( \bar \mu + \delta \mu \right), \quad
       T^0_\alpha = - {1 \over k} \left( \mu + p \right) v_{,\alpha}, 
   \nonumber \\
   & & T^\alpha_\beta = \left( \bar p + \delta p \right) \delta^\alpha_\beta
       + \pi^\alpha_\beta.
   \label{Tab}
\eea
To the background order we have:
\bea
   \mu = {1 \over 2} \sum_J \dot \phi^J \dot \phi_J + V, \quad
       p = {1 \over 2} \sum_J \dot \phi^J \dot \phi_J - V,
   \label{eff-BG}
\eea
where an overdot indicates a time derivative based on $t$.
To the perturbed order we have:
\bea
   & & \delta \mu = \sum_J \left( \dot \phi^J \delta \dot \phi_J
       - \dot \phi^J \dot \phi_J \alpha + V_{,J} \delta \phi^J \right),
   \nonumber \\
   & & \delta p = \sum_J \left( \dot \phi^J \delta \dot \phi_J
       - \dot \phi^J \dot \phi_J \alpha - V_{,J} \delta \phi^J \right),
   \nonumber \\
   & & (\mu + p) v = {k \over a} \sum_J \dot \phi^J \delta \phi_J, \quad
       \pi^\alpha_\beta = 0.
   \label{eff-pert}
\eea

The equations for the background are: 
\bea
   & & H^2 = {8 \pi G \over 3} \left( {1 \over 2} \sum_J \dot \phi^J \dot \phi_J
       + V \right) - {K \over a^2}, 
   \nonumber \\
   & & \dot H = - 4 \pi G \sum_J \dot \phi^J \dot \phi_J + {K \over a^2}, 
   \nonumber \\
   & & \ddot \phi^I + 3 H \dot \phi^I + V^{,I} = 0,
   \label{BG}
\eea
where $K$ is the sign of the spatial curvature, and $H \equiv \dot a/a$.
The first two equations follow from eq. (\ref{GFE}),
and the last one follows from eq. (\ref{EOM}).

Now, we present a complete set of equations describing 
the scalar-type perturbation without fixing the temporal gauge condition,
i.e., in the gauge-ready form:
\bea
   & & \kappa \equiv 3 \left( - \dot \varphi + H \alpha \right)
          + {k^2 \over a^2} \chi,
   \label{G1} \\
   & & - {k^2 - 3K \over a^2} \varphi + H \kappa
       = - 4 \pi G \delta \mu,
   \label{G2} \\
   & & \kappa - {k^2 - 3 K \over a^2} \chi
       = 12 \pi G ( \mu + p ) {a \over k} v,
   \label{G3} \\
   & & \left( \nabla^\alpha \nabla_\beta - {1 \over 3} \delta^\alpha_\beta
       \Delta \right) \left( \dot \chi + H \chi - \alpha - \varphi \right)
       = \pi^\alpha_\beta,
   \label{G4} \\
   & & \dot \kappa + 2 H \kappa
       + \left( 3 \dot H
       - {k^2 \over a^2} \right) \alpha
       = 4 \pi G \left( \delta \mu + 3 \delta p \right),
   \label{G5} \\
   & & \delta \ddot \phi^I + 3 H \delta \dot \phi^I
       + {k^2 \over a^2} \delta \phi^I
       + \sum_J V^{,I}_{\;\;\;\; J} \delta \phi^J
   \nonumber \\
   & & \quad
       = \dot \phi^I \left( \kappa + \dot \alpha \right)
       + \left( 2 \ddot \phi^I + 3 H \dot \phi^I \right) \alpha,
   \label{G6}
\eea
where $\chi \equiv a ( \beta + a \dot \gamma )$ and
$k$ is the comoving wavenumber;
$\nabla_\alpha$ and $\Delta$ are the covariant derivative and 
the Laplacian based on $g_{\alpha\beta}^{(3)}$.
Equations (\ref{G1}-\ref{G6}) are:
the definition of $\kappa$, 
ADM energy constraint ($G^0_0$ component of the field equation),
momentum constraint ($G^0_\alpha$ component), ADM propagation
($G^\alpha_\beta - {1 \over 3} \delta^\alpha_\beta G^\gamma_\gamma$ component),
Raychaudhuri equation ($G^\gamma_\gamma - G^0_0$ component),
and the scalar fields equations of motion, respectively.

In the following we briefly summarize the gauge-ready strategy 
suggested and elaborated in \cite{Bardeen-1988,Hwang-PRW-1991}.
Under the gauge transformation $\tilde x^a = x^a + \xi^a$
with $\xi^t \equiv a \xi^0$ ($0 = \eta$) the perturbed
metric and fluid quantities change as \cite{Hwang-PRW-1991}:
\bea
   & & \tilde \alpha = \alpha - \dot \xi^t, \quad
       \tilde \varphi = \varphi - H \xi^t, \quad
       \tilde \kappa = \kappa + \left( 3 \dot H + {\Delta \over a^2}
       \right) \xi^t,
   \nonumber \\
   & & \tilde \chi = \chi - \xi^t, \quad
       \tilde v = v - {k \over a} \xi^t, \quad
       \delta \tilde \phi^I = \delta \phi^I - \dot \phi^I \xi^t.
   \label{GT}
\eea
As the temporal gauge fixing condition we can impose one condition 
in any of these temporally gauge dependent variables:
$\alpha \equiv 0$ (synchronous gauge),
$\varphi \equiv 0$ (uniform-curvature gauge),
$\kappa \equiv 0$ (uniform-expansion gauge),
$\chi \equiv 0$ (zero-shear gauge),
$v/k \equiv 0$ (comoving gauge),
$\delta \phi^I \equiv 0$ (uniform-$\phi^I$ gauge), etc.
By examining these we notice that,
except for the synchronous gauge condition (which fixes $\alpha = 0$), 
each of the gauge conditions fixes the temporal gauge mode completely.
Thus, a variable in such a gauge condition uniquely corresponds
to a gauge-invariant combination which combines the variable concerned
and the variable used in the gauge condition.
One can recognize the following combinations are gauge-invariant.
\bea
   & & \delta \phi^I_\varphi \equiv \delta \phi^I
       - {\dot \phi^I \over H} \varphi
       \equiv - {\dot \phi^I \over H} \varphi_{\delta \phi^I}, \quad
       \varphi_v \equiv \varphi - {aH \over k} v, 
   \nonumber \\
   & & \varphi_\chi \equiv \varphi - H \chi \equiv - H \chi_\varphi, \quad
       v_\chi \equiv v - {k \over a} \chi, 
   \label{GI}
\eea
etc.
The gauge-invariant combination $\delta \phi_\varphi^I$, for example,
is equivalent to $\delta \phi^I$ in the uniform-curvature gauge which takes
$\varphi \equiv 0$ as the gauge condition, etc.
In this way, we can systematically construct various gauge-invariant
combinations for a given variable.
Since we can make many gauge-invariant combinations even for a given
variable, this way of writing the gauge-invariant combination will turn out
to be convenient practically.
Generally, we do not know the suitable gauge condition {\it a priori}.
The proposal made in \cite{Bardeen-1988,Hwang-PRW-1991} is that we 
write the set of equation without fixing the (temporal) gauge condition 
and arrange the equations so that we can implement easily various 
fundamental gauge conditions:
eqs. (\ref{G1}-\ref{G6}) are arranged accordingly.
We termed it a {\it gauge-ready method}.

\section{Multiple scalar fields}
                                                     \label{sec:MSFs}

The uniform-curvature gauge is convenient for handling $\delta \phi^I$.
Equivalently, we can use the corresponding gauge-invariant combination
$\delta \phi^I_\varphi$ introduced in eq. (\ref{GI}).
Assuming a flat ($K = 0$) background, from eqs. (\ref{G6},\ref{G1}-\ref{G3}) 
we can derive
\bea
   & & \delta \ddot \phi^I_\varphi + 3 H \delta \dot \phi^I_\varphi
       + {k^2 \over a^2} \delta \phi^I_\varphi
   \nonumber \\
   & & \quad
       = - \sum_J \left[ V^{,I}_{\;\;\;\; J} - {8 \pi G \over a^3}
       \left( {a^3 \over H} \dot \phi^I \dot \phi_J \right)^\cdot
       \right] \delta \phi^J_\varphi.
   \label{MSFs-UCG-eq}
\eea
This equation was first derived in \cite{Sasaki-Stewart-1996}.
Compared with $\delta \phi^I$ equation in the uniform-curvature gauge in
eq. (\ref{delta-phi-I-eq}) the ones in the other gauge conditions are more 
complicated \cite{MSF-1994}.
The terms inside parenthesis of the RHS of this equation can be written as
\bea
   V^{,I}_{\;\;\;\; J} + {(8 \pi G)^2 \over H^2} V \dot \phi^I \dot \phi_J 
       + {8 \pi G \over H} \left( V_{,J} \dot \phi^I  
       + V^{,I} \dot \phi_J \right).
\eea
Thus, the scalar field fluctuations are generally coupled to each other
as long as we have $V \neq 0$.

We introduce following gauge-invariant combinations
\bea
   & & \delta \phi_{IJ} \equiv 
       {\delta \phi^I \over \dot \phi^I} - {\delta \phi^J \over \dot \phi^J}, 
   \label{delta-phi_IJ}
\eea
which represent the {\it isocurvature modes}.
If we decompose the effective fluid quantities in 
eqs. (\ref{eff-BG},\ref{eff-pert}) in terms of the sum over individual ones
as $\delta \mu \equiv \sum_J \delta \mu_J$
and $\mu + p = \sum_J (\mu_J + p_J)$ we can show
$S_{IJ} = a^3 ( \delta \phi_{IJ}/a^3)^\cdot$ where
$S_{IJ} \equiv \delta \mu_I / ( \mu_I + p_I) - \delta \mu_J / (\mu_J + p_J)$
is the well known isocurvature perturbation \cite{Kodama-Sasaki-1984}.
{}From eq. (\ref{MSFs-UCG-eq}) we can derive
\bea
   & & {H^2 \over a^3 \dot \phi^I \dot \phi_I} 
       \left( {a^3 \dot \phi^I \dot \phi_I
       \over H^2} \dot \varphi_{\delta \phi^I} \right)^\cdot + {k^2 \over a^2} 
       \varphi_{\delta \phi^I}
   \nonumber \\
   & & \quad
       = - H \sum_J \left[ V^{,I}_{\;\;\;\; J}
       - {8 \pi G \over a^3} \left( {a^3 \over H} \dot \phi^I \dot \phi_J
       \right)^\cdot \right] {\dot \phi^J \over \dot \phi^I} \delta \phi_{IJ}.
   \label{delta-phi-I-eq}
\eea
{}From eqs. (\ref{GI},\ref{eff-pert},\ref{BG}) we have
\bea
   \varphi_v = - {H \over \mu + p} \sum_J \dot \phi_J \delta \phi^J_\varphi,
   \label{varphi_v-eff}
\eea
and from eqs. (\ref{G1},\ref{G3},\ref{varphi_v-eff}) we have
$\varphi_{v} = (H^2/\dot H) \alpha_\varphi$.
Thus, using eq. (\ref{G2}) we can show
\bea
   \dot \varphi_{v} = {k^2 \over a^2} {H \over \dot H} \varphi_\chi
       + {2 H \over (\mu + p)^2} \sum_{J,K}
       V_{,J} \dot \phi^J \dot \phi^K \dot \phi_K \delta \phi_{JK}.
   \label{dot-varphi-eq} 
\eea
{}Finally, we have
\bea
   & & {H^2 \over a^3 \dot H} \left( {a^3 \dot H \over H^2}
       \dot \varphi_{v} \right)^\cdot + {k^2 \over a^2} \varphi_{v}
   \nonumber \\
   & & \quad
       = {2 H^2 \over a^3 ( \mu + p ) }
       \left[ {a^3 \over H ( \mu + p )} \sum_{J,K} V_{,J} \dot \phi^J 
       \dot \phi^K \dot \phi_K \delta \phi_{JK} \right]^\cdot.
   \label{ddot-varphi-eq} 
\eea
In order to derive eq. (\ref{ddot-varphi-eq}) it is useful to have the 
following relation 
\bea
   \varphi_v = \varphi_\chi - {H \over \dot H}
       \left( \dot \varphi_\chi + H \varphi_\chi \right),
   \label{varphi_v-varphi_chi}
\eea
which follows from eqs. (\ref{G1}-\ref{G4}) with eq. (\ref{eff-pert}).
The gauge-invariant combination $\varphi_v$ represents the {\it adiabatic mode}.

Notice that the last terms in eqs. (\ref{dot-varphi-eq},\ref{ddot-varphi-eq})
vanish for the following situations.

(i) A single component case.

(ii) Another trivial case is the one with $V = {\rm constant}$.

(iii) $V_{,I} \propto \dot \phi_I$ for all components.
This applies if we have $3 H \dot \phi^I + V^{;I} = 0$ which implies
$\ddot \phi^I = 0$ in eq. (\ref{BG}); meanwhile 
the slow-roll conditions assume $\ddot \phi^I \ll 3 H \dot \phi^I$.
Thus, in the large-scale limit, ignoring the Laplacian term,
we have a general integral form solution
\bea
   \varphi_{v} ({k}, t) = C ({k}) - D({k})
       \int^t {H^2 \over a^3 \sum_J \dot \phi^J \dot \phi_J} dt,
   \label{MSFs-varphi-LS-sol}
\eea
where $C$ and $D$ are integration constants indicating the coefficients of
relatively growing and decaying solutions, respectively.

(iv) $\delta \phi_{IJ} = 0$.
This case is more restrictive compared with (iii);
the RHS of eq. (\ref{delta-phi-I-eq}) vanishes as well.
The solution in eq.  (\ref{MSFs-varphi-LS-sol}) remains valid, and additionally,
from eqs. (\ref{varphi_v-eff},\ref{delta-phi_IJ}) we can show
$\varphi_{\delta \phi^I} = - H {\delta \phi^I_\varphi / \dot \phi^I} 
 = \varphi_v$.

Thus, under the conditions (i)-(iv)
$\varphi_v$ is {\it conserved} in the large-scale limit even in the 
multi-component situation, see eq. (\ref{MSFs-varphi-LS-sol}).
We note that the gauge-invariant combination $\varphi_v$ which shows
fundamental importance in handling adiabatic perturbation 
was first introduced by Lukash in 1980 \cite{Lukash-1980}

In order to make eq. (\ref{ddot-varphi-eq}) complete we need 
equations for $\delta \phi_{IJ}$.
{}From eq. (\ref{G6},\ref{eff-pert},\ref{G1}-\ref{G3})
we can derive
\bea
   & & {1 \over a^3 \dot \phi_I \dot \phi_J} \left( a^3 \dot \phi_I \dot \phi_J
       \delta \dot \phi_{IJ} \right)^\cdot
   \nonumber \\
   & & \quad
       - \left( {V_{,I} \over \dot \phi_I} - {V_{,J} \over \dot \phi_J} \right)
       \sum_K {\dot \phi^K \dot \phi_K \over \mu + p}
       \left( \delta \dot \phi_{IK} + \delta \dot \phi_{JK} \right)
   \nonumber \\
   & & \quad
       + \left( - 3 \dot H + {k^2 \over a^2} \right) \delta \phi_{IJ}
   \nonumber \\
   & & \quad
       - \sum_K \left( V^{,I}_{\;\;\;\; K} {\dot \phi^K \over \dot \phi^I}
       \delta \phi_{IK}
       - V^{,J}_{\;\;\;\; K} {\dot \phi^K \over \dot \phi^J}
       \delta \phi_{JK} \right)
   \nonumber \\
   & & \quad
       = - 2 \left( {V_{,I} \over \dot \phi_I}
       - {V_{,J} \over \dot \phi_J} \right) 
       {k^2 \over a^2 \dot H} \varphi_\chi
   \nonumber \\
   & & \quad
       = \left( {V_{,I} \over \dot \phi_I} - {V_{,J} \over \dot \phi_J} \right) 
       \left[ - {2 \over H} \dot \varphi_v
       + 4 \sum_{K,L} {V_{,K} \dot \phi^K \dot \phi^L \dot \phi_L
       \over (\mu + p)^2} \delta \phi_{KL} \right],
   \nonumber \\
   \label{ddot-phi_IJ-eq}
\eea
where we used eq. (\ref{dot-varphi-eq}) in the last step.
Notice that only $\dot \varphi_v$ contributes to the isocurvature modes.
Equations (\ref{ddot-varphi-eq},\ref{ddot-phi_IJ-eq})
provide a complete set of equations in terms of 
the adiabatic, $\varphi_v$, and isocurvature, $\delta \phi_{IJ}$,
perturbation variables.
{}For a system of $N$ scalar fields eq. (\ref{MSFs-UCG-eq})
provides a coupled $N$ set of second order differential equations
for $\delta \phi^I$.
Equations (\ref{ddot-varphi-eq},\ref{ddot-phi_IJ-eq})
provide an alternative expression in terms of $\varphi_v$
and $\delta \phi_{IJ}$.

In the case of {\it two} minimally coupled scalar fields,
using $\varphi_v$ and $\delta \phi_{12}$,
eqs. (\ref{ddot-varphi-eq},\ref{ddot-phi_IJ-eq}) lead to
\bea
   & & {H^2 \over a^3 (\mu + p)} \left[ {a^3 (\mu + p) \over H^2}
       \dot \varphi_v \right]^\cdot + {k^2 \over a^2} \varphi_v
   \nonumber \\
   & & \quad
       = {2 H^2 \over a^3 ( \mu + p ) } \left[ {a^3 \dot \phi_1 \dot \phi_2
       ( V_{,1} \dot \phi_2 - V_{,2} \dot \phi_1 ) 
       \over H ( \mu + p )} \delta \phi_{12} \right]^\cdot,
   \label{Phi-eq-2} \\
   & & { \mu + p \over a^3 \dot \phi_1^2 \dot \phi_2^2 }
       \left( {a^3 \dot \phi_1^2 \dot \phi_2^2 \over \mu + p } 
       \delta \dot \phi_{12} \right)^\cdot
       + \Bigg[ - 3 \dot H
       - V_{,12} { \mu + p \over \dot \phi_1 \dot \phi_2 }
   \nonumber \\
   & & \quad
       - 4 \left( {V_{,1} \dot \phi_2 - V_{,2} \dot \phi_1 
       \over \mu + p} \right)^2 
       + {k^2 \over a^2} \Bigg] \delta \phi_{12}
   \nonumber \\
   & & \quad
       = - {2 \over H} \left( {V_{,1} \over \dot \phi_1} 
       - {V_{,2} \over \dot \phi_2} \right) \dot \varphi_v.
   \label{Psi-eq-2}
\eea
Notice that the adiabatic and isocurvature modes are coupled
generally even in the large-scale limit.
Recent study of the two-field system can be found in 
\cite{Gordon-etal-2000}.

\section{Slow-roll situation}
                                                    \label{sec:slow-roll}

We have shown in the paragraph surrounding eq. (\ref{MSFs-varphi-LS-sol})
that under the slow-roll condition, i.e., for $V_{,I} = - 3 H \dot \phi_I$
thus $\ddot \phi_I = 0$, the adiabatic mode decouples from the 
isocurvature modes;
we emphasize again that this condition differs from the ordinary slow-roll 
condition $V_{,I} \simeq - 3 H \dot \phi_I$ used in the inflation models.
Under such a condition the RHS of eq. (\ref{ddot-phi_IJ-eq}) vanishes,
thus the isocurvature modes {\it decouple} from the adiabatic one, as well.
Notice, however, that equations for $\delta \phi_\varphi^I$ in
eq. (\ref{MSFs-UCG-eq}) are still generally coupled.
In such a case eqs. (\ref{ddot-varphi-eq},\ref{ddot-phi_IJ-eq}) become
\bea
   & & {H^2 \over a^3} \left( {a^3 \over H^2}
       \dot \varphi_{v} \right)^\cdot + {k^2 \over a^2} \varphi_{v} = 0,
   \label{SL-varphi_v-eq} \\
   & & {1 \over a^3} \left( a^3 \delta \dot \phi_{IJ} \right)^\cdot
       + \left( - 3 \dot H + {k^2 \over a^2} \right) \delta \phi_{IJ}
   \nonumber \\
   & & \quad
       - \sum_K \left( V^{,I}_{\;\;\;\; K} {\dot \phi^K \over \dot \phi^I}
       \delta \phi_{IK} - V^{,J}_{\;\;\;\; K} {\dot \phi^K \over \dot \phi^J}
       \delta \phi_{JK} \right) = 0.
   \label{SL-phi_IJ-eq}
\eea
In the large-scale limit, we have the solution for an adiabatic mode
\bea
   \varphi_v ({\bf x}, t) = C ({\bf x}) - \tilde D({\bf x})
       \int_0^t {H^2 \over a^3} dt,
   \label{SL-sols-multi1} 
\eea
which is consistent with eq. (\ref{MSFs-varphi-LS-sol}) because
$\sum_J \dot \phi^J \dot \phi_J = $ constant in the slow-roll limit.
Therefore, under the slow-roll conditions, the adiabatic and 
isocurvature modes decouple from each other,
and $\varphi_v$ is temporally conserved in the large-scale limit
for an arbitrary potential. 

We can derive the solution of eq. (\ref{SL-phi_IJ-eq}) for the following cases.

(i) $V(\phi^1, \dots, \phi^N) = \sum_K V_K (\phi^K)$:
In such a case, the term with summation in eq. (\ref{SL-phi_IJ-eq}) vanishes,
and the equation can be written as
\bea
   {1 \over a^3 H^2} \left[ a^3 H^2 \left( {\delta \phi_{IJ} / H} \right)^\cdot 
       \right]^\cdot + {k^2 \over a^2} {\delta \phi_{IJ} / H} = 0,
\eea
thus in the large-scale limit we have a general solution
\bea
   {1 \over H} \delta \phi_{IJ} ({\bf x}, t) = C_{IJ} ({\bf x}) 
       - D_{IJ} ({\bf x}) \int_0^t {dt \over a^3 H^2},
   \label{SL-sols-multi2} 
\eea
where $C_{IJ}$ and $D_{IJ}$ are coefficients of the relatively growing
and decaying solutions.
These behaviors are consistent with the results in 
\cite{Starobinsky-etal} where the nontransient solutions 
in the zero-shear gauge were derived based on some additional conditions;
for translation between the two gauge conditions we can use
\bea
   \varphi_v = \varphi_\chi - {H \over \mu + p}
       \sum_J \dot \phi_J \delta \phi^J_\chi,
   \label{UCG-ZSG}
\eea
which follows from eq. (\ref{varphi_v-eff}), 
and the gauge-invariance of $\delta \phi_{IJ}$. 

(ii) $V(\phi^1, \dots, \phi^N) = \Pi_K V_K (\phi^K)$:
In this case, we can show that the third term with summation 
in eq. (\ref{SL-phi_IJ-eq}) becomes $6 \dot H \delta \phi_{IJ}$,
thus the equation can be written as
\bea
   {H^2 \over a^3} \left[ {a^3 \over H^2} \left( H \delta \phi_{IJ} 
       \right)^\cdot \right]^\cdot + {k^2 \over a^2} H \delta \phi_{IJ} = 0,
\eea
where we ignored quadratic order terms in $\dot H/H^2$.
In the large-scale limit we have a general solution
\bea
   H \delta \phi_{IJ} ({\bf x}, t) = C_{IJ} ({\bf x})
       - D_{IJ} ({\bf x}) \int_0^t {H^2 \over a^3} dt.
   \label{SL-sols-multi3}
\eea
Therefore, when we have potential of additive or multiplicative forms, 
as in (i) or (ii) above, the isocurvature modes decouple from each other, 
and we have $\delta \phi_{IJ} / H$ or $H \delta \phi_{IJ}$
temporally conserved in the large-scale limit, respectively.

(iii) With an arbitrary potential, to the linear order in $\dot H/H^2$, 
the isocurvature modes are generally coupled to each other.
However, since $V_{,IJ} \sim {\cal O} (\dot H)$, if we ignore
linear order terms in $\dot H/H^2$, eq. (\ref{SL-phi_IJ-eq}) becomes
$a^{-3}(a^3 \delta \dot \phi_{IJ})^\cdot +(k^2/a^2) \delta \phi_{IJ} = 0$,
and in the large-scale limit we have
$\delta \phi_{IJ} = C_{IJ} - D_{IJ} \int_0^ta^{-3} dt$, thus
$\delta \phi_{IJ}$'s are conserved.

(iv) In the two component system eq. (\ref{Psi-eq-2}) reduces to
\bea
   & & {1 \over a^3} \left( a^3 \delta \dot \phi_{12} \right)^\cdot
       + \left( - 3 \dot H - V_{,12} { \mu + p \over \dot \phi_1 \dot \phi_2 }
       + {k^2 \over a^2} \right) \delta \phi_{12} = 0.
   \nonumber \\
   \label{Psi-eq-2-two}
\eea
Considering that $V_{,IJ} \sim {\cal O} (\dot H)$ we have
$V_{,12} {\mu + p \over \dot \phi_1 \dot \phi_2} \sim {\cal O} (\dot H)$.
Thus, ignoring quadratic order terms in $\dot H/H^2$,
eq. (\ref{Psi-eq-2-two}) can be written in the following form
\bea
   {1 \over a^3 Q^2} \left[ a^3 Q^2 \left( {\delta \phi_{12} / Q} 
       \right)^\cdot 
       \right]^\cdot + {k^2 \over a^2} {\delta \phi_{12} / Q} = 0, 
\eea
where
\bea
   Q \equiv H e^{ \int V_{,12} {\mu + p \over \dot \phi_1 \dot \phi_2} 
       {dt \over 3H} }.
\eea
In the large-scale limit we have 
\bea
   {1 \over Q} \delta \phi_{12} ({\bf x}, t) 
       = C_{12} ({\bf x}) - D_{12} ({\bf x}) \int_0^t {dt \over a^3 Q^2}.
   \label{SL-sols} 
\eea
Therefore, ignoring the transient solutions we have 
$\varphi_v$ and $\delta \phi_{12}/Q$ remain constant, 
thus $\delta \phi_{12} \propto Q$.
We can show that (i) for $V(\phi_1, \phi_2) = V_1 (\phi_1) + V_2 (\phi_2)$
we have $Q = H$, and (ii) for $V(\phi_1, \phi_2) = V_1 (\phi_1) V_2 (\phi_2)$
we have $Q = 1/H$.
The corresponding solutions in the zero-shear gauge have been investigated
in \cite{Starobinsky-etal}.
In \cite{Starobinsky-etal} 
by imposing additional conditions on perturbed order variables
nontransient parts of the solutions for $\delta \phi^I_\chi$ and $\varphi_\chi$ 
were derived.
The translation into our gauge condition can be made using eq. (\ref{UCG-ZSG}),
and results in \cite{Starobinsky-etal} are consistent
with ours which are represented by the conserved behaviors of 
$\varphi_v$ and $\delta \phi_{12}/Q$.

Aspects of the large scale structure formation with multiple episodes 
of inflation based on the multiple number of minimally coupled scalar field
have been studied in
\cite{Starobinsky-etal,Salopek-etal-1989,Sasaki-Stewart-1996,MSFs}.

\section{Discussons}
                                                \label{sec:Discussion}

In this paper we have investigated aspects of scalar-type 
cosmological perturbation in the context of multiple number of
scalar fields in Einstein gravity.
New results found in this paper are the followings.
We have presented a complete set of scalar-type perturbation 
equation in a gauge-ready form: these are eqs. (\ref{G1}-\ref{G6}).
The conserved behaviors of $\varphi_v$ under certain conditions are
the central theme studied in this paper: see, in particular, the 
paragraph below eq. (\ref{varphi_v-varphi_chi}), and eqs. 
(\ref{MSFs-varphi-LS-sol},\ref{SL-sols-multi1}).
A set of closed form differential equations in terms of the
adiabatic and the isocurvature perturbation variables, $\varphi_v$ and 
$\delta \phi_{IJ}$, is derived in
eqs. (\ref{ddot-varphi-eq},\ref{ddot-phi_IJ-eq}).
The equations in two-component situation are presented
in eqs. (\ref{Phi-eq-2},\ref{Psi-eq-2}).
Under the slow-roll conditions the adiabatic and the isocurvature
modes decouple from each other,
see eqs. (\ref{SL-varphi_v-eq},\ref{SL-phi_IJ-eq}),
and various general solutions are presented below eq. (\ref{SL-phi_IJ-eq}).
In the large-scale limit we have
(i) the adiabatic mode is generally conserved, eq. (\ref{SL-sols-multi1}), 
(ii) for a couple of special potential
the isocurvature modes decouple from each other and are described by
conserved quantities, eq. (\ref{SL-sols-multi2},\ref{SL-sols-multi3}),
(iii) in the two field system, the isocurvature mode is described by
a conserved quantity for the general potential, eq. (\ref{SL-sols}).

A complete set of perturbation equations in the case of 
{\it generalized gravity} with multiple minimally coupled scalar fields 
is presented in Sec. 4.2 of \cite{Hwang-PRW-1991} in the gauge ready form.
Our complete set of equations in the gauge-ready form and other closed form 
equations and the large-scale solutions derived in this paper will be 
useful in handling situations involving 
multiple number of scalar fields in diverse cosmological situations.

\subsection*{Acknowledgments}

HN was supported by KOSEF.
We thank Christopher Gordon, Ewan Stewart, David Wands, and Winfried Zimdahl 
for useful discussions.

\baselineskip 0pt

\end{document}